\documentclass[aps,pra,superscriptaddress,twocolumn]{revtex4-2}
\usepackage{eurosym}
\usepackage{amsfonts}
\usepackage{mathrsfs}
\usepackage{color}
\usepackage{xspace}
\usepackage{longtable}
\usepackage{multirow}
\usepackage{tabularx}
\usepackage{amsmath}
\usepackage{amssymb}
\usepackage{graphicx}
\usepackage{dcolumn}
\usepackage{bm}
\usepackage{siunitx}
\usepackage{float}
\usepackage[colorlinks,urlcolor=blue,citecolor=blue,linkcolor=blue]{hyperref}
\usepackage[normalem]{ulem}
\usepackage{epstopdf}
\usepackage{appendix}
\usepackage[level]{datetime}
\usepackage{verbatimbox}
\usepackage{pifont}
\newcommand{\e}{\mathrm{e}}
\newcommand{\SO}{\mathrm{SO}}
\newcommand{\deltatwo}{\frac{\delta}{2}}
\newcommand{\up}{\uparrow}
\newcommand{\dn}{\downarrow}
\newcommand{\DD}{\mathrm{DD}}
\begin{document}

\title{Density-Dependent Gauge Field with Raman Lattices}

\author{Xiang-Can Cheng}
\author{Zong-Yao Wang}
\author{Jinyi Zhang}
\author{Shuai Chen}
\affiliation{Hefei National Research Center for Physical Sciences at the Microscale and School of Physical Sciences, University of Science and Technology of China, Hefei 230026, China}
\affiliation{Shanghai Research Center for Quantum Sciences and CAS Center for Excellence in Quantum Information and Quantum Physics, University of Science and Technology of China, Shanghai 201315, China}
\affiliation{Hefei National Laboratory, University of Science and Technology of China, Hefei 230088, China}
\author{Xiaotian Nie}
\email{nxtnxt@hfnl.cn}
\affiliation{Hefei National Laboratory, University of Science and Technology of China, Hefei 230088, China}
\date{\today}

\begin{abstract}
The study of the gauge field is an everlasting topic in modern physics. Spin-orbit coupling is a powerful tool in ultracold atomic systems, resulting in an artificial gauge field that can be easily manipulated and observed in a tabletop environment. Combining optical Raman lattices and atom-atom interaction, the artificial gauge field can be made density-dependent. In this work, we propose a straightforward way to engineer one-dimensional density-dependent gauge field in a Bose-Hubbard model in spin-orbit coupled Raman lattices. Next, we study the model from two perspectives: few-body quantum walk dynamics and many-body ground state. In the first perspective, we show that large spin-flipped tunneling can lead to a deep two-body bound state. In the second perspective, mean-field and density matrix renormalization group (DMRG) calculations consistently reveal three different phases, i.e. the Mott insulator phase, the superfluid phase, and the magnetic superfluid phase. Finally, we discuss the experimental protocol with Raman lattices based on existing experimental platforms.
\end{abstract}

\maketitle

\section{Introduction}
Gauge theory is the cornerstone of modern physics. Elementary particles interact with each other through gauge fields, which is at the heart of the standard model \cite{weinbergQuantumTheoryFields1995,gattringerQuantumChromodynamicsLattice2010,yangObservationGaugeInvariance2020}. In high energy physics, gauge fields are governed by physical laws, and the study of gauge fields involves accelerating particles and observe their collision \cite{ellisQCDColliderPhysics1996,suColdAtomParticleCollider2024}. In solid-state systems, artificial gauge fields can be exerted externally \cite{klitzingNewMethodHighAccuracy1980,laughlinQuantizedHallConductivity1981,thoulessQuantizedHallConductance1982} or designed intrinsically \cite{haldaneModelQuantumHall1988,qiTopologicalQuantizationSpin2006,changExperimentalObservationQuantum2013}, which would result in topological bands and lead to novel materials with vast application potential \cite{changColloquiumQuantumAnomalous2023}. Alternatively, gauge fields can be simulated within ultracold atom systems, by manipulating the geometry phase the atoms acquire while interacting with light \cite{dalibardColloquiumArtificialGauge2011,goldmanLightinducedGaugeFields2014,cooperTopologicalBandsUltracold2019}. The simulation of gauge theory in ultracold atom systems has been a fruitful area of study throughout the years, with plenty of proposals \cite{dumGaugeStructuresAtomLaser1996,visserGeometricPotentialsSubrecoil1998,duttaTunnelingDynamicsGauge1999,juzeliunasEffectiveMagneticFields2005,osterlohColdAtomsNonAbelian2005,ruseckasNonAbelianGaugePotentials2005,juzeliunasLightinducedEffectiveMagnetic2006,zhuSpinHallEffects2006,gunterPracticalSchemeLightinduced2009,spielmanRamanProcessesEffective2009,campbellRealisticRashbaDresselhaus2011,andersonSynthetic3DSpinOrbit2012,xuDynamicalGenerationArbitrary2012} and experiments \cite{madisonVortexFormationStirred2000,linBoseEinsteinCondensateUniform2009,linSyntheticMagneticFields2009,linSyntheticElectricForce2011,aidelsburgerRealizationHofstadterHamiltonian2013,atalaDirectMeasurementZak2013,miyakeRealizingHarperHamiltonian2013,parkerDirectObservationEffective2013,struckEngineeringIsingXYSpinmodels2013} to create and observe artificial gauge fields.

In the quantum simulations of artificial gauge fields mentioned above, the gauge fields are static, only providing a background stage for matter's evolution. Recently, the focus of the study has advanced to the simulation of the dynamical gauge field, which is more of an analog to the real gauge fields in the physical world \cite{clarkObservationDensityDependentGauge2018,gorgRealizationDensitydependentPeierls2019,schweizerFloquetApproachZ22019,yangObservationGaugeInvariance2020,frolianRealizing1DTopological2022,yaoDomainwallDynamicsBose2022}. In these systems, atoms (matter) give a back action to the gauge field. Therefore, the gauge field is no longer a background field but a degree of freedom evolving self-consistently. One practical way to create a dynamical gauge field is to make the gauge potential density-dependent using atom-atom interaction. In ultracold atoms, interaction can be added with Feshbach resonances \cite{chinFeshbachResonancesUltracold2010,jurgensenObservationDensityInducedTunneling2014,frolianRealizing1DTopological2022}, or by confining atoms in deep optical lattices \cite{greinerQuantumPhaseTransition2002,schweizerFloquetApproachZ22019,yangObservationGaugeInvariance2020}.

The realization of spin-orbit coupling (SOC) through Raman transitions in ultracold atoms enables new ways to study topological bands and gauge fields \cite{qiTopologicalQuantizationSpin2006,linSpinOrbitcoupledBose2011,zhengCollectivemodesSOCBEC2012,zhengPropertiesBoseGases2013,jiExperimentalDeterminationFinitetemperature2014,zhengFloquettopologicalstates2014,wuRealizationTwodimensionalSpinorbit2016,huangExperimentalRealizationTwodimensional2016,sunHighlyControllableRobust2018,sunUncoverTopologyQuantum2018,zhangTwoLegSSH2017,langNodalBrillouinzone2017,hasanWavePacketDynamics2022,liangRealizationQiWuZhangModel2023}. The first realization of spin-orbit coupled Bose-Einstein condensates \cite{linSpinOrbitcoupledBose2011} is based on the same experimental apparatus that generates artificial gauge fields \cite{linSyntheticMagneticFields2009,linBoseEinsteinCondensateUniform2009,linSyntheticElectricForce2011}. Similar idea was followed to engineer SOC in optical lattices \cite{wuRealizationTwodimensionalSpinorbit2016,sunHighlyControllableRobust2018,wangRealizationIdealWeyl2021}, where retro-reflected laser beams generate standing waves and form optical lattices, as well as Raman couplings. These Raman lattices can flip atom spin while transferring momenta, thus creating SOC. Since the realization of SOC with ultracold atoms, there has been a vast number of theoretical studies that introduce on-site interaction to spin-orbit coupled bosons in lattices \cite{coleBoseHubbardModelsSynthetic2012a,radicExoticQuantumSpin2012,bolukbasiSuperfluidMottinsulatorTransition2014a,hickeyThermalPhaseTransitions2014a,zhaoEvolutionMagneticStructure2014a,zhaoFerromagnetismTwocomponentBoseHubbard2014a,gongDzyaloshinskiiMoriyaInteractionSpiral2015,yanSpinorbitdrivenTransitionsMott2017,xuDensitydependentSpinorbitCoupling2021}. These works investigated the ground state spin texture as well as the phase transitions that was brought about by SOC. Recently, a proposal was made to create a non-Abelian dynamical gauge field and topological superfluids for fermions with optical Raman lattices \cite{zhouNonAbelianDynamicalGauge2023}. However, the density-dependent gauge field generated by SOC in a Bose-Hubbard landscape has yet been covered.

In this work, we introduce on-site interaction to a one-dimensional Raman lattice model similar to those in references \textcite{wuRealizationTwodimensionalSpinorbit2016,sunHighlyControllableRobust2018}. We show that a density-dependent gauge field emerges spontaneously in low energy from this spin-orbit coupled Bose-Hubbard model. Next, we study the model from two different perspectives: few-body dynamics and many-body ground state. For the first perspective, we investigate the two-body quantum walk under the density-dependent spin-conserved and -flipped tunneling. Within different parameter regimes and initial states, the time evolution and density-density correlation functions show different behaviors, such as asymmetric tunneling, confinement, etc. Crucially, via two spin-flip steps, two atoms of opposite spin can move through each other. In the \(t_\SO\) dominant regime, the dynamics favor this process over moving apart, resulting in confinement. Energy spectra in the quasi-momentum space reveal that the bound states become deeper and overlap more with the initial states with larger \(t_\SO\). In the second perspective, we study the many-body ground state phase diagram with a clustered Gutzwiller method \cite{natuStaticDynamicProperties2015,yanDynamicsDisorderedStates2017,yanEquilibrationDynamicsStrongly2017} and collaborate it with DMRG calculations. A Mott insulator (MI) phase, a superfluid (SF) phase, and 
 a magnetic superfluid (MSF) phase have been found. Finally, we give an experimental protocol based on Raman lattices and additional deep optical lattices to realize the SOC and the Hubbard interaction.

\section{Model}
We consider a one-dimensional Bose-Hubbard model with SOC. The Hamiltonian can be written as \cite{wuRealizationTwodimensionalSpinorbit2016,wangDiracRashbaWeyltype2018}
\begin{equation}
    \begin{aligned}
        \hat{H}_\mathrm{Hubbard}=&\hat{H}_0+\hat{H}_\mathrm{int}-\sum_j\mu\hat{n}_j,\\
        \hat{H}_0=&\sum_j\hat{\mathbf{\Psi}}_j^\dagger\frac{\delta}{2}\sigma_z\hat{\mathbf{\Psi}}_j+\sum_{j}(\hat{\mathbf{\Psi}}_{j+1}^\dagger T_\mathrm{SOC}\hat{\mathbf{\Psi}}_j+\mathrm{H.c.}),\\
        \hat{H}_\mathrm{int}=&\sum_j[\frac{U_{\uparrow\uparrow}}{2}\hat{n}_{j\up}(\hat{n}_{j\up}-1)\\
        &+\frac{U_{\dn\dn}}{2}\hat{n}_{j\dn}(\hat{n}_{j\dn}-1)
        +U_{\up\dn}\hat{n}_{j\up}\hat{n}_{j\dn}],
    \label{BHSOC}
    \end{aligned}
\end{equation}
where \(T_\mathrm{SOC}=t_0\sigma_z-i t_\SO\sigma_y\) is the tunneling matrix, \(t_0\) \((t_\SO)\) is the spin-conserved (spin-flipped) tunneling strength, \(U_{\sigma\sigma'}(\sigma=\up,\dn)\) are the spin dependent on-site interaction strengths, \(\mu\) is the chemical potential, \(\hat{\mathbf{\Psi}}^\dagger_j=(\hat{a}^\dagger_{j\uparrow},\hat{a}^\dagger_{j\downarrow})\) denotes the creation operators of the spinful bosons, \(\hat{n}_{j\sigma}=\hat{a}^\dagger_{j\sigma}\hat{a}_{j\sigma}\) is the number operator, \(\hat{n}_j=\hat{n}_{j\up}+\hat{n}_{j\dn}\), \(\sigma_{y,z}\) are the Pauli matrices. \(\delta/2\) is the Zeeman splitting, corresponds to the two-photon detuning of the Raman coupling. The \(t_0\sigma_z\) term indicates that the spin-conserved tunneling coefficients for spin-\(\up\) atoms and spin-\(\dn\) atoms have opposite phase, which comes from the relative spatial configuration of the lattice and the Raman potentials, namely, the periodicity of the Raman potentials is one-half of the lattice period. This property is essential, and in a 2D non-interacting system leads to a quantum anomalous Hall model driven by SOC \cite{wuRealizationTwodimensionalSpinorbit2016}. In a Hubbard model, the on-site interaction energy depends on the occupation. Thus, generally speaking, large interactions can strongly suppress the tunneling process. However, in our system, the spin-flipped tunneling can be restored by a Raman process with two-photon detuning \(\delta\sim U\), so that the overhead energy from a spin flip can be compensated by the interaction energy difference before and after the tunneling \cite{jurgensenObservationDensityInducedTunneling2014,xuCorrelatedSpinflipTunneling2018}. Therefore, whether the spin-conserved or -flipped tunneling can happen depends on the density difference between the two sites. 

\begin{figure}[ht]
    \centering
    \includegraphics[width=0.45\textwidth]{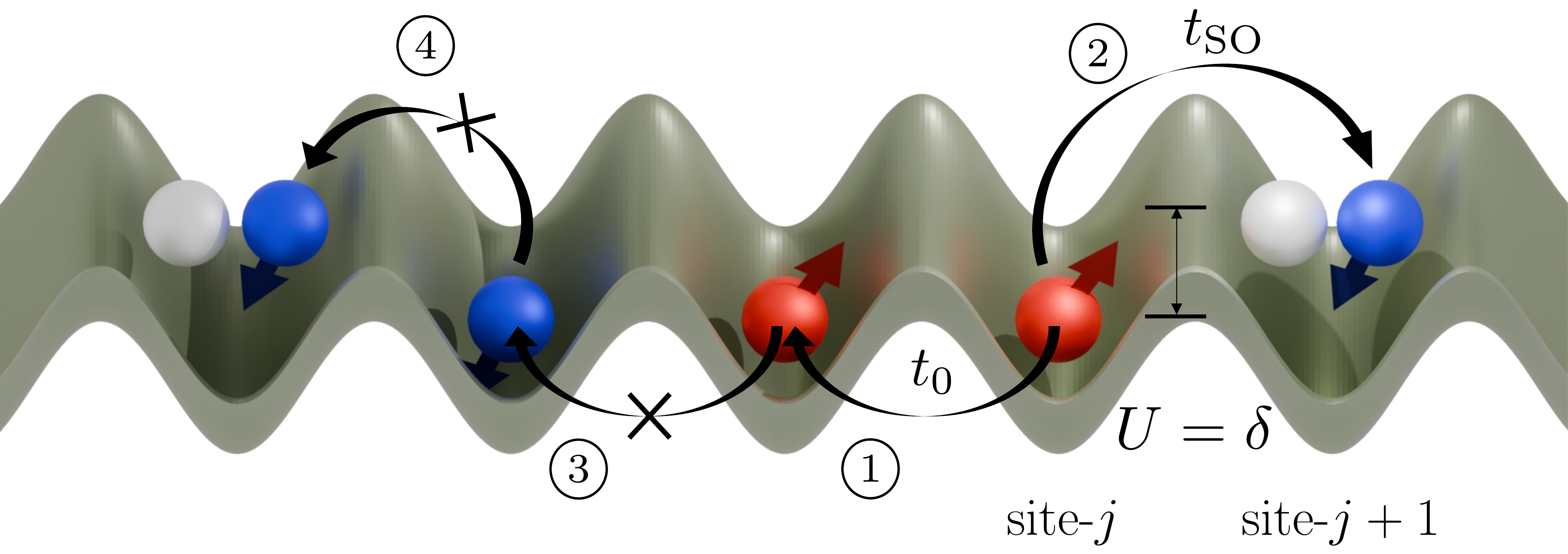}
    \caption{Physical picture of the density-dependent tunneling. Red balls denote spin-\(\up\) atoms, while blue ones are spin-\(\dn\) atoms. The white ball represents an arbitrary spin. The spin-\(\up\) state and the spin-\(\dn\) state have an energy difference \(\delta\) due to the Zeeman splitting. This energy difference can be compensated by the interaction energy change \(U\) when atoms tunnel to neighboring sites. For examples, tunnelings labeled as \normalsize{\textcircled{\scriptsize{1}}} to \normalsize{\textcircled{\scriptsize{4}}} help to illustrate the density-dependent tunneling (see main text).}
    \label{fig1}
\end{figure}

FIG. \ref{fig1} gives the physical picture of the density-dependent tunneling. The tunneling can be illustrated with four different processes. FIG. \ref{fig1} \normalsize{\textcircled{\scriptsize{1}}} shows a spin-\(\up\) atom tunneling to a neighboring empty site with its spin conserved. There is no energy cost whatsoever, therefore allowed in the low-energy limit. This also applies to spin-\(\dn\) atoms. FIG. \ref{fig1} \normalsize{\textcircled{\scriptsize{2}}} is a \(\up\) atom tunneling to a neighboring single-occupied site and flipping spin. The interaction energy increases by \(U\), which can be compensated by the Zeeman energy change \(\delta\) if \(U=\delta\). Thus, this process is also permitted. So is its reverse process, namely a spin-\(\dn\) atom in a double-occupied site tunneling to a neighboring empty site and flipping spin. These are the only allowed dynamics. FIG. \ref{fig1} \normalsize{\textcircled{\scriptsize{3}}} shows a spin-\(\up\) atom tunneling to a neighboring empty site and flipping spin. This will result in an energy change by \(\delta\), meaning the process is off-resonance, and prohibited in the low energy limit. Finally, a spin-\(\dn\) atom tunneling to a neighboring single-occupied site without flipping spin, shown in FIG. \ref{fig1} \normalsize{\textcircled{\scriptsize{4}}}, will lead to a change in interaction energy by \(U\), which also makes the process off-resonance. 

Under a unitary transformation and rotating-frame approximation \footnote{See Appendix \ref{DerivationAppendix} for a detailed derivation of the gauge transformation.}, the Hamiltonian turns into 
\begin{equation}
\begin{aligned}
    \hat{H}_{\DD}&=\sum_{j}(\hat{\mathbf{\Psi}}_{j+1}^\dagger T_{\DD}\hat{\mathbf{\Psi}}_j+\mathrm{H.c.})\\
    &+\sum_j[\frac{\tilde U_{\up\up}}{2}\hat{n}_{j\up}(\hat{n}_{j\up}-1)\\
    &+\frac{\tilde U_{\dn\dn}}{2}\hat{n}_{j\dn}(\hat{n}_{j\dn}-1)
    +\tilde U_{\up\dn}\hat{n}_{j\up}\hat{n}_{j\dn}
    -\mu \hat{n}_j],
\end{aligned}
\end{equation}
in which the tunneling matrix becomes density-dependent, 
\begin{equation}
    T_{\mathrm{\DD}}=
        \begin{pmatrix}
            \hat{P}^0 t_0&-\hat{P}^{-1} t_{\SO}\\
            \hat{P}^{+1} t_{\SO}&-\hat{P}^0 t_0
        \end{pmatrix}.
    \label{DDTunneling}
\end{equation}
\(\hat{P}^{\Delta n}=\sum_{n}\hat{P}^{n+\Delta n}_{j+1}\hat{P}^{n}_{j}\) is the projection operator, projecting the system into the subspace where atom numbers differ in \(\Delta n\) between neighboring sites. \(\tilde U_{\sigma\sigma'}=U_{\sigma\sigma'}-\delta\) are the effective interaction strengths. This model requires \(U_{\sigma\sigma'}\) and \(\delta\) to be much greater than \(t_0\), \(t_\SO\) and \(\mu\), while \(\tilde U_{\sigma\sigma'}\) should be comparable to \(t_0\), \(t_\SO\) and \(\mu\). In this system, whether a tunneling process can happen depends on the density difference between the two sites. Note that the projection operator sits between the annihilation operator \(\Psi_j\) and the creation operator \(\Psi^\dagger_{j+1}\), which means the density difference of projection is calculated after an atom has been annihilated in site-\(j\), but before it is created in site-\(j+1\). For example, consider the tunneling \normalsize{\textcircled{\scriptsize{2}}} in FIG. \ref{fig1}, before the tunneling the two sites have the same density. After the spin-\(\up\) atom hops out of site-\(j\) and before it hops into site-\(j+1\), the occupation difference \(\Delta n\) is \(1\). Thus \(\hat{P}^{+1}\) applies and only a spin-flipped hopping can happen. 

\section{Perspective 1: Few-body dynamics}
\begin{figure}[ht]
    \centering
    \includegraphics[width=0.45\textwidth]{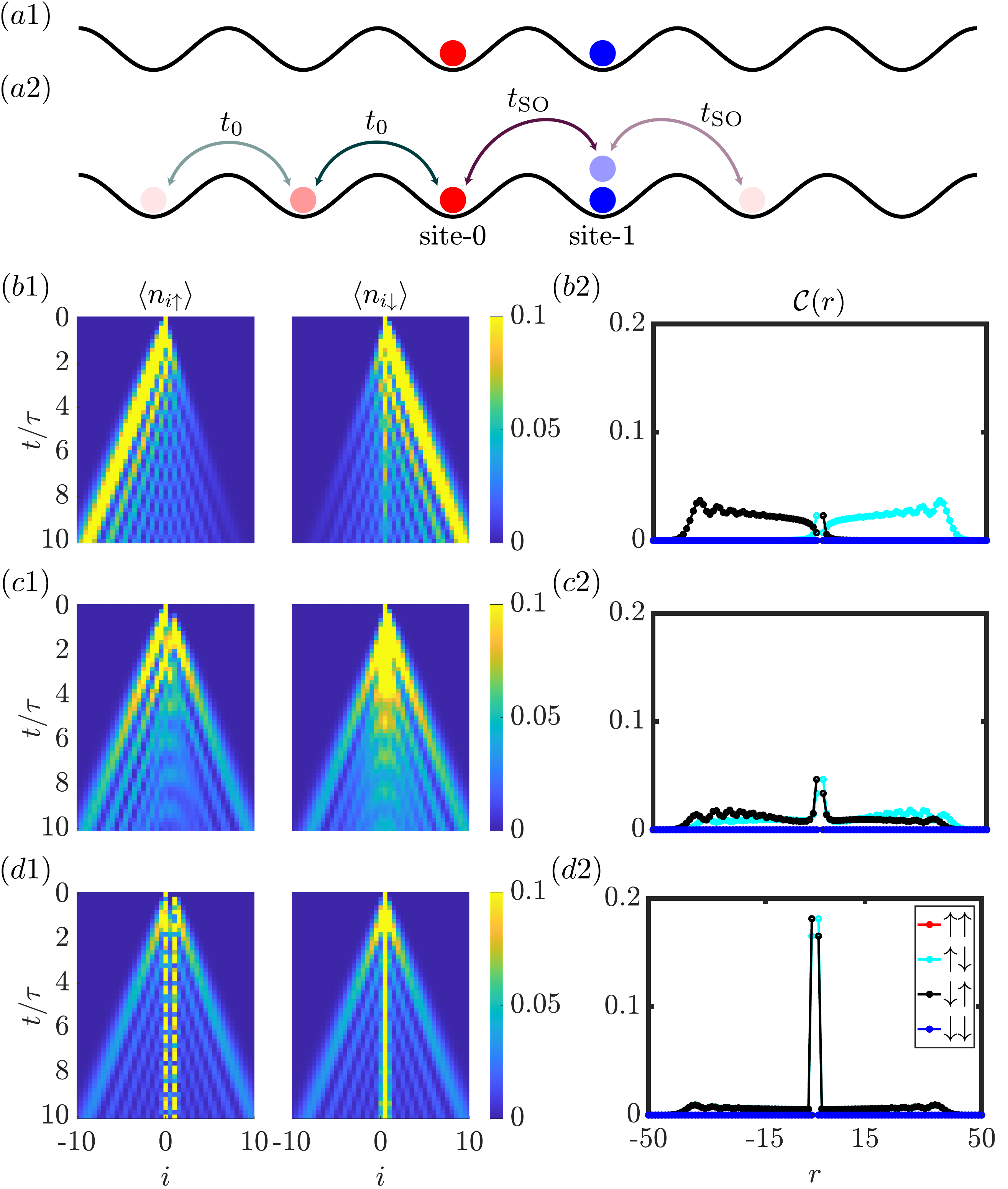}
    \caption{Time evolution of initial state \(|\dots,0,0,\up,\dn,0,0,\dots\rangle\). (a1) Sketch of the initial state. Red circle represents a spin-\(\up\) atom, blue circle represents a spin-\(\dn\) atom. (a2) Sketch of some typical density-dependent tunneling processes. The spin-\(\up\) atom can either tunnel to the left via spin-conserved tunnelings, or tunnel through the spin-\(\dn\) atom to the rightmost side via two spin-flipped tunnelings. (b1-d1) Density evolution of spin-\(\up\) and spin-\(\dn\) at \(t_\SO/t_0=0.1, 1, 10\). (b2-d2) Correlation function over relative distance \(r\) for \(t_\SO/t_0=0.1, 1, 10\). The correlation function is the snapshot at \(t=10\tau\).}
    \label{0u1d}
\end{figure}

We first consider the two-body quantum walk, to study the dynamics of the atoms influenced by the density-dependent gauge field. In this section, we set \(U_{\up\up}=U_{\dn\dn}=U_{\up\dn}=\delta\) for simplicity, i.e. \(\tilde{U}_{\sigma\sigma'}=0\), and focus only on the effect of the density-dependent tunneling. The chemical potential \(\mu\) is also set to zero. We fix the spin-conserved tunneling strength \(t_0\) and vary the spin-flipped tunneling strength \(t_\SO/t_0\) in 0.1, 1, and 10. The Hamiltonian becomes
\begin{equation}
    \hat{H}_{\DD}=\sum_{j}(\hat{\mathbf{\Psi}}_{j+1}^\dagger T_{\DD}\hat{\mathbf{\Psi}}_j+\mathrm{H.c.}).
\end{equation}
We start the system at initial site configuration \(|\dots,0,0,\up,\dn,0,0,\dots\rangle\), i.e. two atoms of opposite spin starting from neighboring sites [Fig. \ref{0u1d} (a1)], and calculate the dynamics along a chain from \(t=0\) to \(t=10\tau\), \(\tau=1/t_0\). In FIG. \ref{0u1d} we present the density distribution of the two spins at each time, as well as the two-body correlation function in relative positions \(\mathcal{C}_{\sigma,\sigma'}(r)=\sum_i\langle n_{i\sigma}n_{(i+r)\sigma'}\rangle|_{t=10\tau}\) (\(\sigma\),\(\sigma'\)=\(\up\) or \(\dn\)) at the end of the dynamics \cite{kwanRealizationOnedimensionalAnyons2024}.

Under the density-dependent tunneling, the two atoms need two spin-flip processes to go through one another [Fig. \ref{0u1d} (a2)]. For \(t_\SO/t_0=0.1\), the spin-conserved tunneling is dominant. Therefore, each spin can propagate to their corresponding half of the lattice, but can hardly cross each other. The correlation shows that the two atoms have a high probability spacing far apart, with the spin-\(\dn\) atom at the right. If we increase \(t_\SO/t_0\) to 1, the spin-flipped tunneling will be enhanced. Different spins can move through each other more easily, so the asymmetry is reduced. The correlation is becoming more symmetric as well, and a peak starts to show up at near zero distance, indicating a shallow bound state. Finally, if we set \(t_\SO/t_0\) to \(10\), the spin-flipped tunneling becomes the dominant process. The evolution shows much more obvious confinement behavior. However, there is still a small chance that the atoms can be separated far apart due to a minor spin-conserved hopping. In the correlation, the confinement peak is very prominent. The two spins are more likely to be bound together. The small plateau corresponds to the free propagation dynamics due to the spin-conserved tunneling.

\begin{figure*}[ht]
    \centering
    \includegraphics[width=0.8\textwidth]{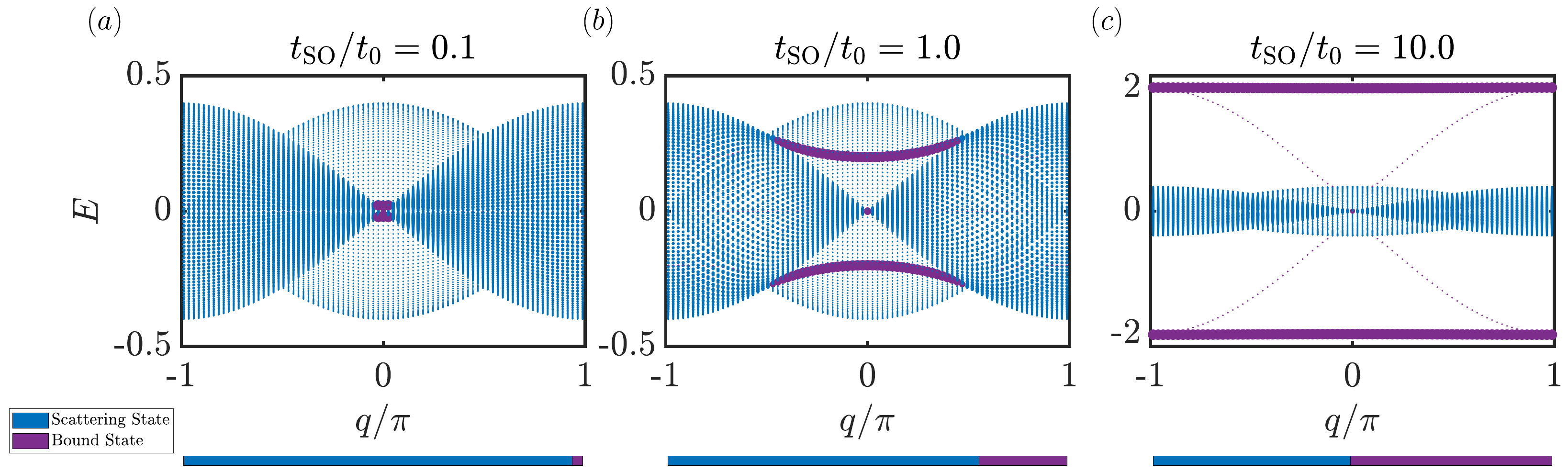}
    \caption{Energy spectra in the total quasi-momentum space. (a-c) Energy spectra for \(t_\SO/t_0=0.1,1,10\). Each dot represents an eigenstate. Blue dots are scattering states, purple dots are bound states, and the size of a dot is proportional to the overlap between the initial state and this eigenstate plus a constant shift, to make all eigenstates visible. Below each plot, a bar chart shows the sum of overlap on all bound states against all scattering states. For \(t_\SO/t_0=0.1,1,10\), the total overlap on all bound states are \(0.026, 0.220, 0.505\).}
    \label{BoundState}
\end{figure*}
We obtain the energy spectra labeled by total quasi-momentum to study the bound states, shown in FIG. \ref{BoundState}. Each dot represents an eigenstate. Blue dots are scattering states and purple dots are bound states \footnote{See Appendix \ref{BoundStateAppendix} for details in distinguishing the bound states.}. The size of the dot denotes the overlap between the initial state and the eigenstate. All three spectra consist of a continuum of states, majorly from scattering states made of two far-separated free atoms. The bound states are in the continuum in small \(t_\SO\) regime and become isolated outside of the continuum when \(t_\SO\) is dominant. For \(t_\SO/t_0=0.1\), the initial state mostly overlaps with the scattering states. Increasing \(t_\SO\), the overlap with the bound states grows. For \(t_\SO/t_0=10\), the overlap between the initial state and the bound states is about \(0.505\), which agrees with the dynamics observed before.

\section{Perspective 2: Many-body ground state}

\subsection{Mean-Field}
Next, we come to study the many-body physics and calculate the mean-field phase diagram of the density-dependent Hamiltonian. In this section, we set \(\tilde{U}_{\up\up}=\tilde{U}_{\up\dn}=\tilde{U}_{\dn\dn}-\Delta \tilde U=\tilde{U}\) as an energy scale. Here we add a spin imbalance \(\Delta \tilde U=0.1\tilde U\) to the interaction to explicitly break the \(\mathrm{SU}(2)\) symmetry of interaction. This can avoid possible degeneracy between doubly occupied states. The density-dependent Hubbard Hamiltonian becomes:
\begin{equation}
\begin{aligned}
    \hat{H}_{\DD}&=\sum_{j}(\hat{\mathbf{\Psi}}_{j+1}^\dagger T_{\DD}\hat{\mathbf{\Psi}}_j+\mathrm{H.c.})+\sum_j[\frac{\tilde U}{2}\hat{n}_j(\hat{n}_j-1)\\
    &+\frac{\Delta \tilde U}{2}\hat{n}_{j\downarrow}(\hat{n}_{j\downarrow}-1)-\mu \hat{n}_j].
\end{aligned}
\end{equation}
We employ a 2-site clustered Gutzwiller variational wavefunction \cite{luhmannClusterGutzwillerMethod2013} to find the ground state of the system:
\begin{equation}                
    |\mathrm{G}\rangle=\prod_{\mathrm{2-site\atop clusters}}\sum_{l_{1\up},l_{1\dn},\atop l_{2\up},l_{2\dn}}f_{l_{1\up},l_{1\dn},l_{2\up},l_{2\dn}}|l_{1\up},l_{1\dn},l_{2\up},l_{2\dn}\rangle,
\end{equation}
where \(l_{j\up(\dn)}\) is the number of atoms at site-\(j\) with spin-\(\up\) (\(\dn\)). We set the cut-off of atom number of each spin at \(l_{j\up(\dn)}\leq l_{\mathrm{max}}=2\). The energy of the trial state is \(E=\langle \mathrm{G}|\hat{H}|\mathrm{G}\rangle\). The variational parameters 
\(f_{l_{1\up},l_{1\dn},l_{2\up},l_{2\dn}}\) 
can be determined by minimizing \(E\), with the normalization condition \(\sum|f_{l_{1\uparrow},l_{1\downarrow},l_{2\uparrow},l_{2\downarrow}}|^2=1\).

\begin{figure*}[ht]
    \centering
    \includegraphics[width=0.8\textwidth]{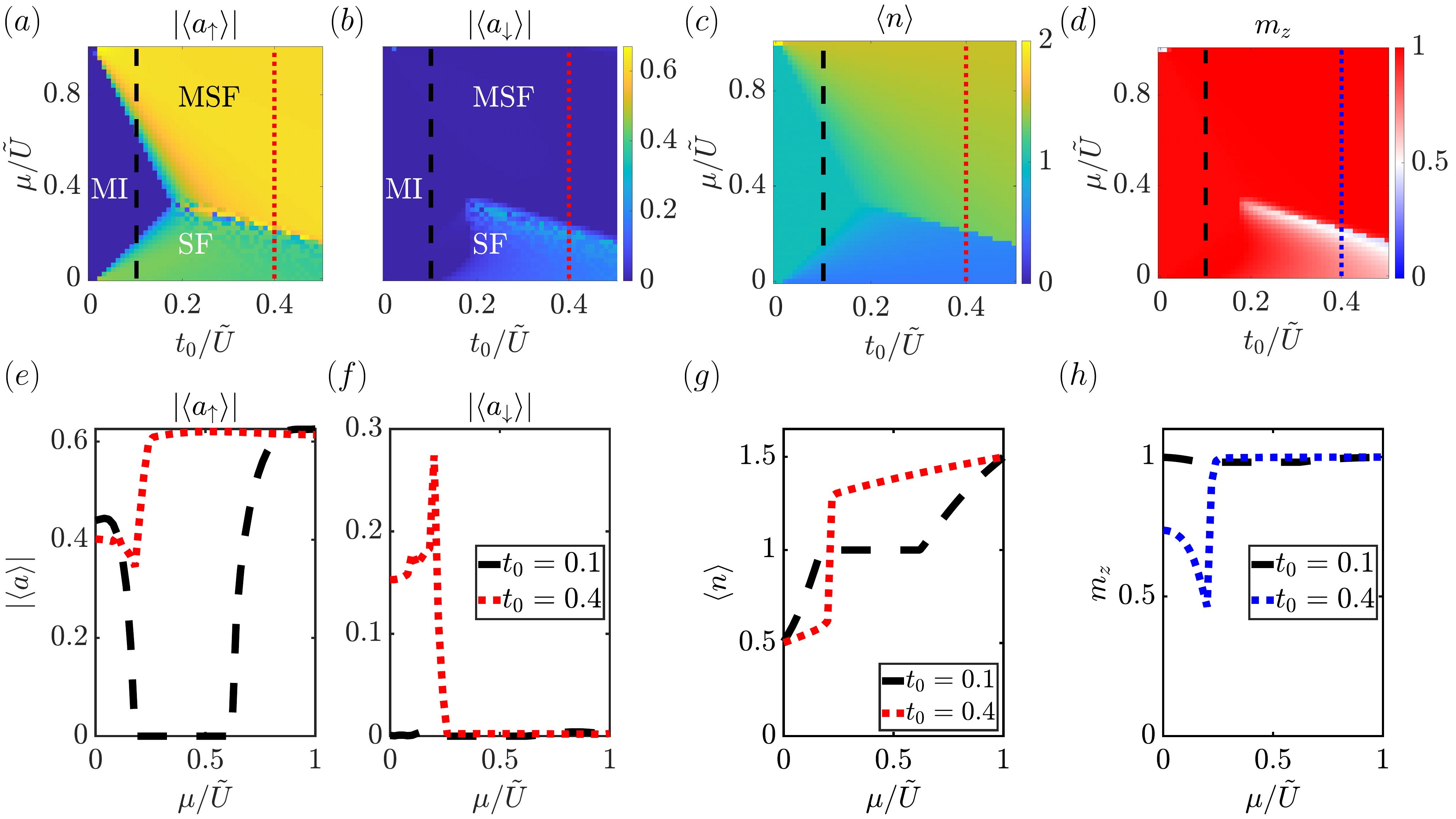}
    \caption{Mean-field phase diagrams. (a-b) The magnitude of the superfluid order parameter \(\langle a_{\sigma}\rangle\). (c) Phase diagram of atom number expectation value \(\langle n\rangle\). (d) Phase diagram of normalized spin polarization. (e-f) Magnitude of \(\langle a\rangle\) along two vertical lines, \(t_0=0.1\tilde U\) and \(t_0=0.4\tilde U\). (g) \(\langle n\rangle\) along two vertical lines. (h) Spin polarization \(m_z\) along two different vertical lines. }
    \label{PDs}
\end{figure*}

Our results are shown in FIG. \ref{PDs}. We found three distinct phases, a Mott insulator (MI) phase, a superfluid (SF) phase, and a magnetic superfluid (MSF) phase which is fully polarized. The superfluid order parameter is defined as \(\langle a_{j\sigma}\rangle=\langle \mathrm{G}|a_{j\sigma}|\mathrm{G}\rangle\). For \(a_{1\up}\), \(\langle a_{1\up}\rangle=\sum_{l_{1\up},l_{1\dn},l_{2\up},l_{2\dn}}f^*_{l_{1\up}-1,l_{1\dn},l_{2\up},l_{2\dn}}f_{l_{1\up},l_{1\dn},l_{2\up},l_{2\dn}}\sqrt{l_{1\up}}\), etc. Our calculation shows that the superfluid order parameter \(|\langle a_{j\sigma}\rangle|\) and other observable expectation values are uniform across the two sites. Therefore, we only show the results of the first site of the cluster and drop the site index \(j\). FIG. \ref{PDs} (a-d) show the phase diagrams of \(t_0/\tilde U\) versus \(\mu/\tilde U\) at \(t_\SO=0.1\tilde U\). The color in (a-b) represents the magnitude of the superfluid order parameter \(\langle a_{\up(\dn)}\rangle\). In the MI phase, both \(\langle a_{\up}\rangle\) and \(\langle a_{\dn}\rangle\) vanish. In the SF phase, both \(\langle a_{\up}\rangle\) and \(\langle a_{\dn}\rangle\) have non-zero values. In the MSF phase, \(\langle a_{\dn}\rangle\) is nearly zero, but \(\langle a_{\up}\rangle\) is non-zero. From the vertical lines of \(t_0=0.1\) and \(t_0=0.4\) (FIG. \ref{PDs} (e-f)), we can see the phase transitions between the MI phase and the SF/MSF phase are second-order phase transitions. The SF-MSF phase transition is first-order. 

In FIG. \ref{PDs} (d,h) we plot the spin polarization \(m_z=(\langle n_{\up}\rangle-\langle n_{\dn}\rangle)/(\langle n_{\up}\rangle+\langle n_{\dn}\rangle)\). The spin in the MSF phase is fully polarized to spin-\(\up\), there is no occupation at spin-\(\dn\). Meanwhile, the SF phase is only partially polarized. 

\subsection{DMRG}
\begin{figure*}[ht]
    \centering
    \includegraphics[width=0.8\textwidth]{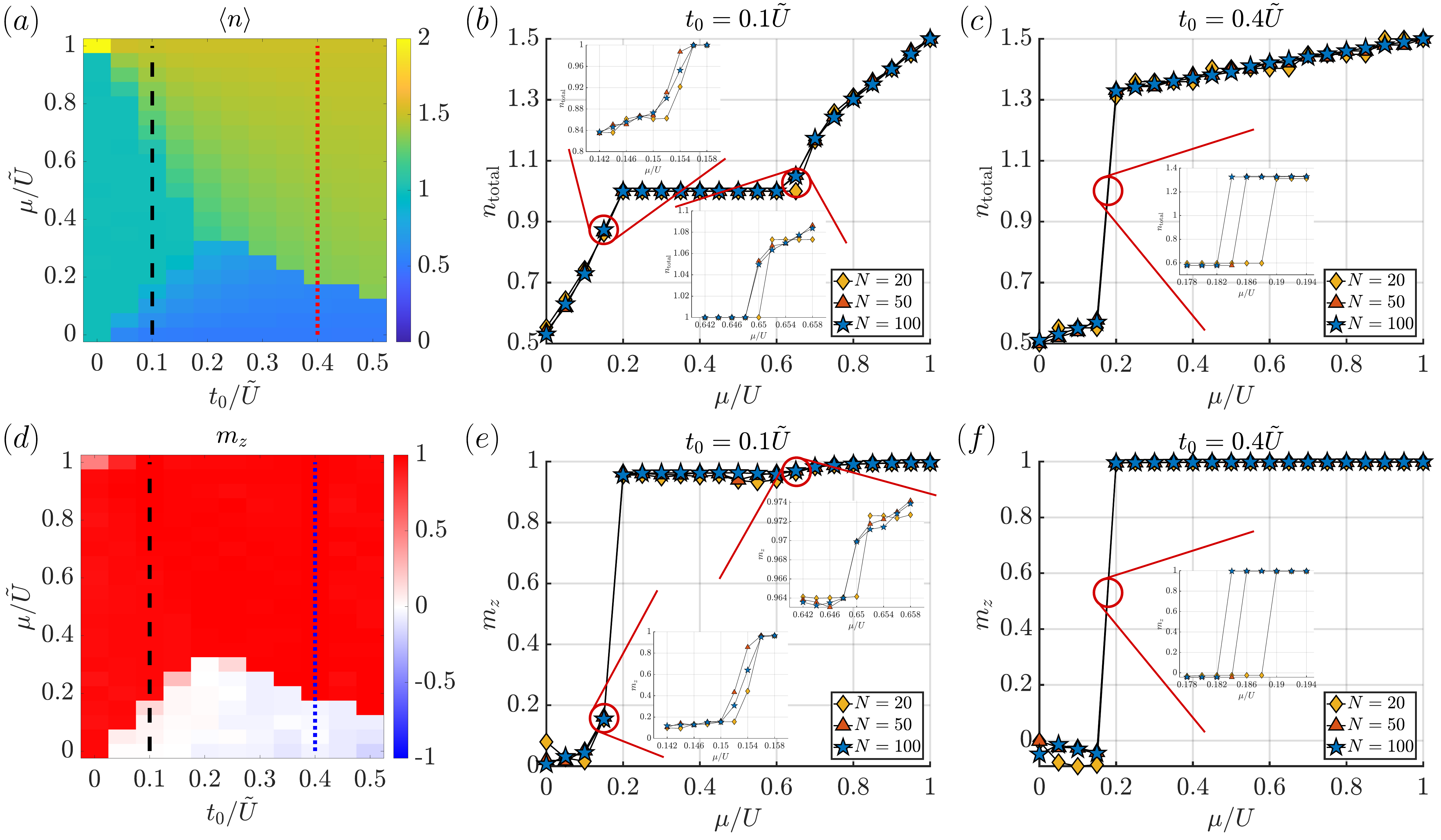}
    \caption{DMRG results. (a) Phase diagram of the expectation value of on-site atom number \(\langle n\rangle\), \(N=100\). (b-c) \(\langle n\rangle\) along two vertical lines, \(t_0=0.1\tilde U\) and \(t_0=0.4\tilde U\). (d) Phase diagram of the expectation value of spin polarization \(\langle m_z\rangle\), \(N=100\). (e-f) \(\langle m_z\rangle\) along two vertical lines, \(t_0=0.1\tilde U\) and \(t_0=0.4\tilde U\). }
    \label{DMRG}
\end{figure*}

Quantum fluctuation becomes more dominant in low-dimensional systems. There are some situations where the mean-field approximation would be drastically influenced or even fail in 1D lattice models. To collaborate with our mean-field results, we perform DMRG calculations with the density-dependent gauge field Hamiltonian. We consider open boundary condition with system size \(N=20,50,100\). The cut-off of atom number of each spin is \(5\), and the maximum bond dimension is 800. In the DMRG method, since the Mermin-Wagner theorem excluded spontaneous breaking of continuous symmetry in 1D quantum systems, the superfluid phases are now distinguished by their compressibility. On the contrary, the Mott insulator phase shows incompressibility (integer filling). Here we plot the expectation value of the on-site atom number, as well as the spin polarization in FIG. \ref{DMRG}. 

FIG. \ref{DMRG} (a), (d) show the phase diagrams of \(\langle n\rangle\) and \(m_z\), which are qualitatively the same as the mean-field results. The contrast between the SF phase and the MSF phase is much greater than the mean-field result: the magnetization is almost \(0\) for SF and \(1\) for MSF. FIG. \ref{DMRG} (b), (e) and (c), (f) are the \(\langle n\rangle\) and \(m_z\) for \(t_0=0.1\tilde U\) and \(t_0=0.4\tilde U\), respectively. Insets are zoom-ins near phase transitions. All of the zoom-ins are plotted in intervals of the same width near the transitions, \(\Delta t_0=0.002\tilde U\). We can see that the phase transitions between SF-MI-MSF along \(t_0=0.1\tilde U\) are continuous, while the phase transition between SF and MSF along \(t_0=0.4\tilde U\) is first-order. The high degree of consistency between the two results indicates that our 2-site clustered Gutzwiller ansatz is capable of capturing the physics of the density-dependent gauge field.

\section{Experimental Protocol}
\begin{figure}[ht]
    \centering
    \includegraphics[width=0.4\textwidth]{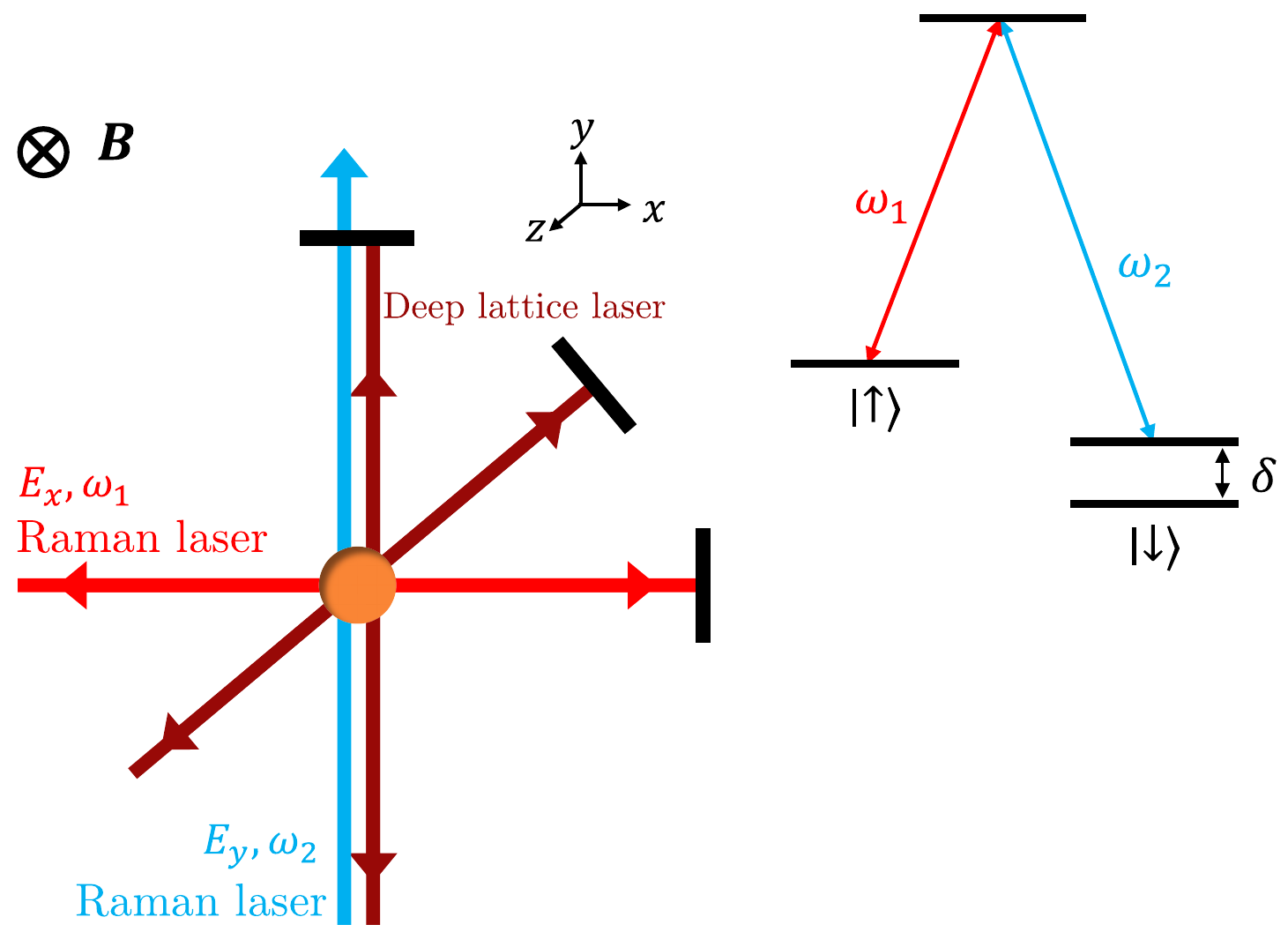}
    \caption{Experimental scheme to implement the density-dependent gauge field. A retro-reflected Raman beam along \(x\) direction and a running beam along \(y\) creates the 1D SOC and Raman lattices. An additional set of deep lattices along \(y\) and \(z\) confines atoms around lattice sites. A filter in the \(y\) direction path retro-reflects the deep lattice beam while the Raman running beam transmits. A bias magnetic field provides the Zeeman splitting. The detuning \(\delta\) can be tuned by setting the frequency of one of the Raman lasers.}
    \label{Exp}
\end{figure}
Our model can be implemented on existing experimental platforms\cite{wuRealizationTwodimensionalSpinorbit2016,sunHighlyControllableRobust2018,zhangSpinorbitCouplingTopological2018} with Raman lattices. A proposed scheme is shown in FIG. \ref{Exp}. A retro-reflected laser beam along the \(x\) direction, together with a running beam with the same wavelength along \(y\) can create the desired 1D Raman lattice and introduce the SOC. The Raman detuning \(\delta\) can be adjusted by setting the Raman laser frequency, and the Raman coupling strength \(t_\SO\) can be tuned by setting the relative phase between Raman lasers. Two additional sets of retro-reflected laser beams along \(y\) and \(z\) directions can create deep optical lattices and localize the atoms around lattice sites, thereby increasing the on-site interaction \(U\), which can be tuned by changing the lattice depths. The density-dependent gauge field emerges when \(U\) and \(\delta\) are much larger than \(t_0\), \(t_\SO\) and \(\mu\), while \(U\sim \delta\). 

As a reference, for Rubidium-87 atoms a \(\SI{787}{nm}\) laser can be chosen to form the Raman lattices, and the deep lattices can be generated from a \(\SI{1064}{nm}\) laser. If the deep lattice depth is set to \(5 E_\mathrm{r} \-- 30 E_\mathrm{r}\), the Raman lattice depth is set to \(3 E_\mathrm{r} \-- 10 E_\mathrm{r}\), and the Raman coupling strength is set to \(2 E_\mathrm{r}\), under the tight-binding approximation the interaction strength and the tunneling strengths can be tuned within \(t_0\approx\SIrange[range-phrase=\--]{70}{400}{Hz}\), \(t_\mathrm{SO}\approx\SIrange[range-phrase=\--]{70}{700}{Hz}\), \(U\approx\SIrange[range-phrase=\--]{1.5}{0.3}{kHz}\), which are experimentally feasible and enough to cover the phase diagram. Chemical potential \(\mu\) can be tuned by changing the atom number. Finally, a quantum gas microscope can be utilized to probe the quantum walk dynamics and the two body correlation function\cite{preissStronglyCorrelatedQuantum2015,kwanRealizationOnedimensionalAnyons2024}. The MI-SF and SF-MSF phase transition can be studied with spin-resolved time of flight (TOF) imaging.

\section{Summary}
In this work, we constructed one-dimensional density-dependent dynamical synthetic gauge field from a spin-orbit coupled Bose-Hubbard model. Starting from the experimentally realized scheme for SOC, we showed that the density-dependent gauge field emerges in low energy when the Raman detuning compensates the on-site interaction and restores the spin-flipped hopping.

To study new effects of density-dependent gauge field, we investigate from two perspectives: few-body dynamics and many-body ground state.
For the few-body perspective, we calculated the quantum walk of two opposite spins. We found that by increasing the spin-flipped tunneling strength \(t_\SO\), two atoms can form a confinement pair, which corresponds to bound states in the total quasi-momentum space. For the many-body perspective we looked into its phase diagram. We obtained the mean-field phase diagram with a clustered Gutzwiller method. We found three distinct phases, the MI phase, the SF phase, and the MSF phase. We found a second-order phase transition between the superfluid phases and the MI phase, and a first-order phase transition between the SF phase and the MSF phase. The mean-field method was collaborated with DMRG calculations.

Last, we gave a proposal to experimentally realize this model with Raman lattices in cold atom systems. We conclude that our prediction can be tested within currently available experimental platforms. We believe that the study of density-dependent dynamical gauge field can be continued along this path to higher dimensions and non-Abelian situations, where more new physics can be found.
\begin{acknowledgments}
We thank Professor Wei Zheng for valuable guidance and Jizhou Wu, Hao-Jie Wu, and Xin-Chi Zhou for insightful discussions. This work was supported by the National Natural Science Foundation of China (Grant No. 12025406). J.Z. acknowledges support from the CAS Talent Introduction Program (Category B) (Grant No. KJ9990007012) and the Fundamental Research Funds for the Central Universities. The numerical calculations in this paper have been done on the supercomputing system in the Supercomputing Center of University of Science and Technology of China. The DMRG calculations are performed with the ITENSOR library\cite{ITensor,ITensorCode}.
\end{acknowledgments}

\appendix

\section{Derivation of the Density-Dependent Gauge Field From the Original Bose-Hubbard Model}\label{DerivationAppendix}
In this appendix, we give a detailed derivation of the unitary transformation that leads to the density-dependent Hamiltonian. Our goal is to formally let the Raman detuning term \(\delta/2\) and the interaction term \(U_{\sigma,\sigma'}\) cancel out, and turn them into density-dependent phases in the tunneling matrix. 

We start with the Bose-Hubbard Hamiltonian \eqref{BHSOC}, and apply the first unitary transformation \(G=\mathrm{e}^{i\deltatwo t\sum_j(\hat{n}_{j\uparrow}-\hat{n}_{j\downarrow})}\). The Hamiltonian reads
\begin{equation}
    \begin{aligned}
        \hat{H}_{\mathrm{GT}}&=G\hat{H}_{\mathrm{Hubbard}}G^\dagger-G(i\partial_t)G^\dagger\\
        &=\sum_{j}(\hat{\Psi}^\dagger_{j+1}T_{\mathrm{GT}}\hat{\Psi}_j+\mathrm{H.c.})+\sum_j[\frac{U_{\up\up}}{2}\hat{n}_{j\up}(\hat{n}_{j\up}-1)\\
        &+\frac{U_{\dn\dn}}{2}\hat{n}_{j\dn}(\hat{n}_{j\dn}-1)+U_{\up\dn}\hat{n}_{j\up}\hat{n}_{j\dn}-\mu\hat{n}_j],
    \end{aligned}
\end{equation}
where the tunneling matrix takes the form
\begin{equation}
    T_{\mathrm{GT}}=t_0\sigma_z-it_{\SO}
    \begin{pmatrix}
        0&-i\e^{i\delta t}\\
        i\e^{-i\delta t}&0
    \end{pmatrix}.
\end{equation}

Here we can see, the Raman detuning \(\delta/2\) turns into a time-dependent phase in the spin-flipped tunneling. We then take the second unitary transformation \(R=\e^{i\frac{\delta}{2} t\sum_j\hat{n}_j(\hat{n}_j-1)}\). The Hamiltonian becomes
\begin{equation}
    \begin{aligned}
        \hat{H}_{\DD}&=R\hat{H}_{\mathrm{GT}}R^\dagger-R(i\partial_t)R^\dagger\\
        &=\sum_{j}(\hat{\Psi}^\dagger_{j+1}T_{\mathrm{DD}}\hat{\Psi}_{j}+\mathrm{H.c.})+\sum_j[\frac{\tilde U_{\up\up}}{2}\hat{n}_{j\up}(\hat{n}_{j\up}-1)\\
        &+\frac{\tilde U_{\dn\dn}}{2}\hat{n}_{j\dn}(\hat{n}_{j\dn}-1)+\tilde{U}_{\up\dn}\hat{n}_{j\up}\hat{n}_{j\dn}-\mu\hat{n}_j],
    \end{aligned}
    \label{HDD}
\end{equation}
where the tunneling matrix is
\begin{equation}
    T_{\mathrm{DD}}=
    \begin{pmatrix}
        t_0\e^{i\delta t\Delta \hat{n}_{j}}&-t_{\SO}\e^{i\delta t(\Delta \hat{n}_{j}+1)}\\
        t_{\SO}\e^{i\delta t(\Delta \hat{n}_{j}-1)}&-t_0\e^{i\delta t\Delta \hat{n}_{j}}
    \end{pmatrix},
    \label{DDTunnelingFull}
\end{equation}
\(\Delta\hat n_{j}=\hat{n}_{j+1}-\hat{n}_{j}\) is the atom number difference operator between neighboring sites. Here, the tunneling matrix has density-dependent phases, and \(\tilde{U}_{\sigma,\sigma'}=U_{\sigma,\sigma'}-\delta\) is the residual effective interaction. Then, we adopt the rotating-wave approximation to drop the time-dependent oscillating terms, which means projecting the Hamiltonian to the subspace of \(\Delta {n}=0,1,-1\), for different elements of the tunneling matrix. Labeling the atom number projection operator on site-\(j\) \(\hat{P}^{n}_{j}\) , and \(\hat{P}^{\Delta n}=\sum_{n}\hat{P}^{n+\Delta n}_{j+1}\hat{P}^{n}_{j}\), we finally arrive at the tunneling matrix in \eqref{DDTunneling}.

\section{Distinguishing the Bound States}\label{BoundStateAppendix}
To distinguish the bound states from the scattering states, we check the two body density correlation at a large distance, \(\langle n_0n_{L/2}\rangle\), which represents the largest distance in the chain under periodic boundary condition. Naturally, we assert that this correlation should almost vanish for bound states since particles in bound states cannot be separated at a far distance. 

\begin{figure*}[ht]
    \centering
    \includegraphics[width=0.8\textwidth]{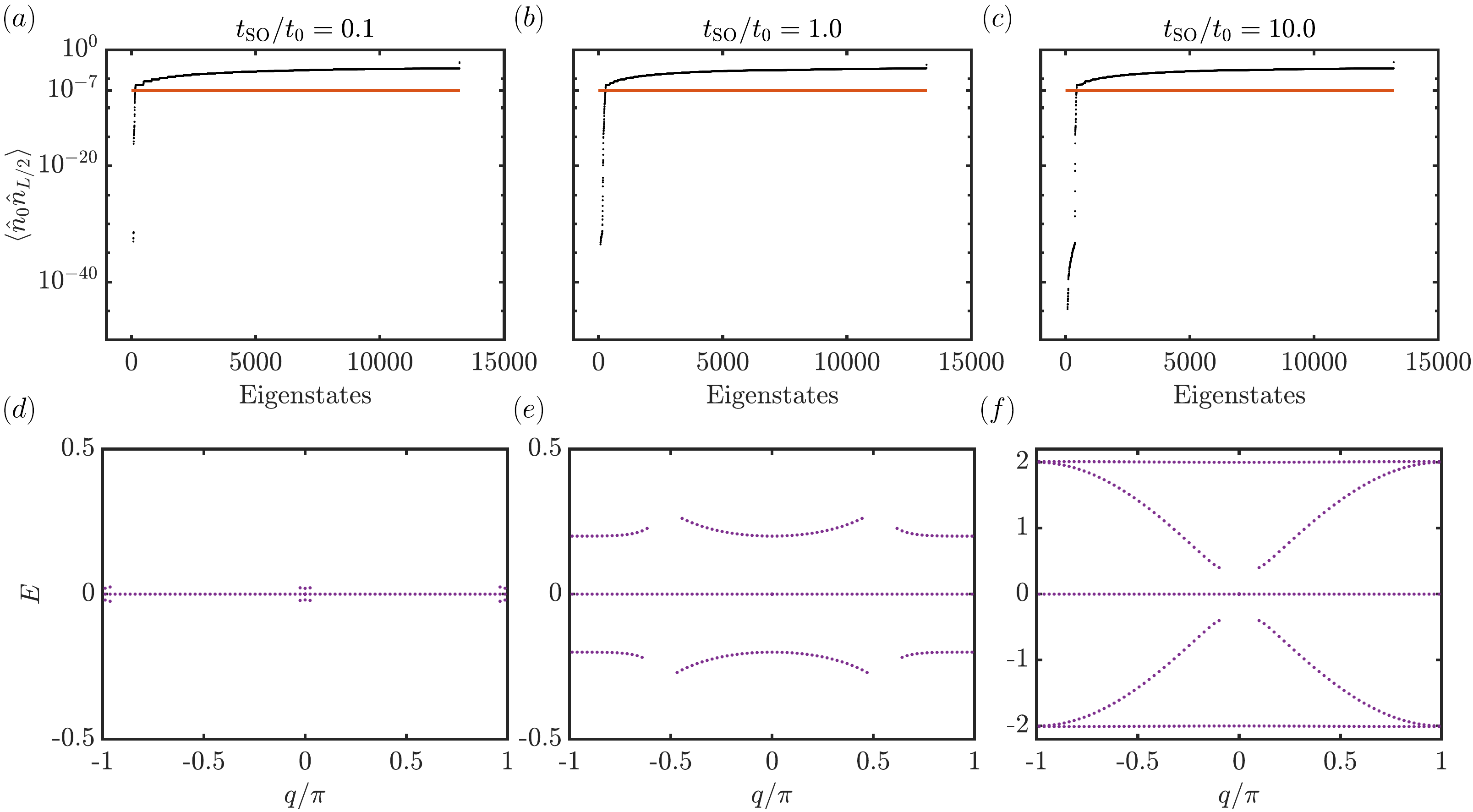}
    \caption{Distinguishing the bound states. (a-c) Far away correlation \(\langle n_0n_{L/2}\rangle\) in log scale, for \(t_\SO/t_0=0.1,1,10\). The orange horizontal line shows the threshold for bound state \SI{1e-7}{}. (d-f) Bound states in the energy spectra.}
    \label{CorrFarFig}
\end{figure*}
The correlations of all eigenstates in the momentum space for each \(t_\SO\) are sorted and plotted in FIG. \ref{CorrFarFig} (a-c). A clear transition shows up at around \(\SI{1e-7}{}\), which we choose as the threshold for bound states. (We have checked that a small deviation in the threshold does not change the conclusion.) The bound states are plotted in FIG. \ref{CorrFarFig} (d-f). Among them, there is a set of trivial bound states which corresponds to double \(\up\) atoms occupying one site, \(|\dots,0,0,\up\up,0,0,\dots\rangle\). These states are not coupled to any basis by the Hamiltonian. Therefore, they form a zero-energy band with vanishing group velocity.

\section{Quantum Walk of the Non-Interacting System}
In this section, we show the quantum walk dynamics of a non-interacting system without the density dependence in the tunneling matrices. The Hamiltonian is
\begin{equation}
\begin{aligned}
    \hat{H}_{\mathrm{NI}}&=\sum_j\hat{\mathbf{\Psi}}_{j+1}^\dagger T_{\mathrm{NI}}\hat{\mathbf{\Psi}}_j+\mathrm{H.c.},\\
    T_{\mathrm{\mathrm{NI}}}&=
        \begin{pmatrix}
            t_0&-t_{\SO}\\
            t_{\SO}&-t_0
        \end{pmatrix}.
\end{aligned}
\end{equation}

\begin{figure}[ht]
    \centering
    \includegraphics[width=0.45\textwidth]{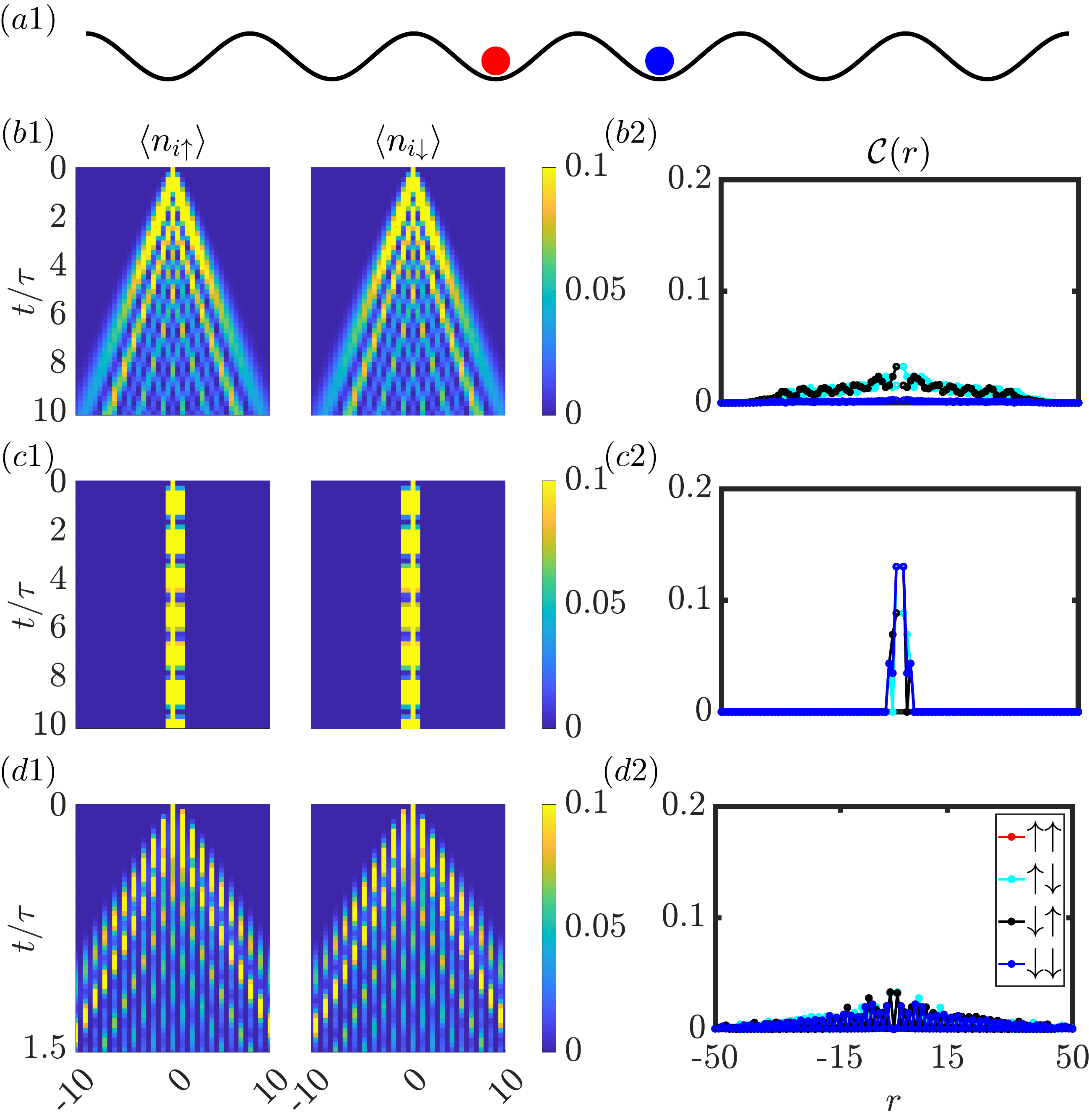}
    \caption{Time evolution of initial state \(|\dots,0,0,\up,\dn,0,0,\dots\rangle\) for the non-interacting system. (a) Sketch of the initial state. Red circle represents a spin-\(\up\) atom, blue circle represents a spin-\(\dn\) atom. (b1-d1) Density evolution of spin-\(\up\) and spin-\(\dn\) at \(t_\SO/t_0=0.1, 1, 10\). The cut-off time for \(t_\SO=10\) is chosen at \(t=1.5\tau\) since the atoms are approaching the boundaries. (b2-d2) Correlation function over relative distance \(r\).}
    \label{NI0u1d}
\end{figure}

Quantum walk evolution for \(|\dots,0,0,\up,\dn,0,0,\dots\rangle\) is shown in figures \ref{NI0u1d}. We can see that the dynamics are much different without the density dependence, since it is just wave packet diffusion of two non-interacting atoms. 

Consequently, some density-dependence-induced dynamics like the asymmetric evolution and confinement are not present in the density-independent dynamics, for the atoms have no trouble moving through each other now. Note that the non-propagating dynamics for \(t_\SO/t_0=1\) is the coincidental result of a flat band for these particular parameters.

\end{document}